\documentclass[twocolumn,showpacs,prb]{revtex4}
\usepackage{epsfig}

\newcommand{\be}{\begin{equation}}
\newcommand{\ee}{\end{equation}}
\newcommand{\beqn}{\begin{eqnarray}}
\newcommand{\eeqn}{\end{eqnarray}}
\newcommand{\nsw}{N_{\mathrm{sweep}}}
\newcommand{\nsa}{N_{\mathrm{samp}}}
\newcommand{\varql}{{\mathrm Var}(\ql)}
\newcommand{\ql}{q_{l}}
\newcommand{\qbar}{{\bar q}}

\newcommand{\ml}{\mu_{l}}

\begin{document}

\title{Monte Carlo simulations of the four-dimensional $XY$ spin glass at low
temperatures}
\author{Helmut G. Katzgraber}
\email{dummkopf@physics.ucsc.edu}
\altaffiliation{Present address: Department of Physics, University of
California, Davis, California 95616}

\author{A. P. Young}
\email{peter@bartok.ucsc.edu}
\homepage{http://bartok.ucsc.edu/peter}
\altaffiliation{Present address:
Department of Theoretical Physics, 1, Keble Road, Oxford OX1 3NP,
England}
\affiliation{Department of Physics,
University of California,
Santa Cruz, California 95064}

\date{\today}

\begin{abstract}
We report on the results for simulations of the four-dimensional
$XY$ spin glass using the parallel tempering Monte Carlo method at low
temperatures for moderate sizes.
Our results are qualitatively consistent with earlier work on the
three-dimensional gauge glass as well as three- and four-dimensional
Edwards-Anderson Ising spin glass. An extrapolation of our results would
indicate that large-scale excitations
cost only a finite amount of energy in the thermodynamic limit.
The surface of these excitations may be fractal, although we cannot
rule out a scenario compatible with replica symmetry breaking in which
the surface of low-energy large-scale excitations is space filling.
\end{abstract}

\pacs{75.50.Lk, 75.40.Mg, 05.50.+q}
\maketitle

\section{Introduction}
\label{introduction}

There has been an ongoing controversy regarding the spin-glass phase.
There are two main theories: the ``droplet picture''
(DP) by Fisher and Huse\cite{fisher:87} and the replica symmetry breaking
picture (RSB) by Parisi.\cite{parisi:79,mezard:87}
While RSB follows the exact solution of the Sherrington-Kirkpatrick model
and predicts that excitations which involve a finite fraction of the spins
cost a finite energy in the thermodynamic limit, the droplet picture states 
that a cluster of spins of size $l$ costs an energy proportional to
$l^{\theta}$, 
where $\theta$ is positive. It follows that in the thermodynamic limit,
excitations that flip a finite cluster of spins cost an infinite energy.
In addition, the DP states that these excitations are fractal with a fractal 
dimension $d_s < d$, where $d$ is
the space dimension, whereas in RSB these excitations
are space filling,\cite{marinari:00} i.e., $d_s = d$.

Krzakala and Martin,\cite{krzakala:00} as well 
as Palassini and Young\cite{palassini:00a} (referred to as KMPY) found, on the
basis of numerical results on small systems with Ising symmetry,
that an intermediate picture may be present: while
the surface of large-scale excitations appears to be fractal, only a finite 
amount of energy is needed to excite them in the thermodynamic limit. 
In the context of their work, it is necessary to introduce two exponents,
$\theta$ and $\theta'$, where $L^\theta$ is the typical energy for an
excitation induced by a change in boundary conditions in a system of linear size
$L$, and $L^{\theta^\prime}$
describes the energy of thermally excited system-size clusters.
Subsequently, similar results were found for the three-dimensional gauge
glass,\cite{katzgraber:01a} which has a continuous symmetry but is known to
have a finite $T_c$. 

The differences between DP and RSB can be
quantified by studying the distribution
\cite{marinari:00,katzgraber:01,reger:90,marinari:98a,zuliani:99} $P(q)$
of the spin overlap $q$ defined in Eq.~(\ref{qmunu-xy}) below.
For finite
systems, the DP predicts two peaks at $\pm q_{\rm EA}$, where $q_{\rm EA}$ is the
Edwards-Anderson order parameter, as well as a tail down to $q = 0$ that
vanishes in the thermodynamic limit like\cite{bray:86,moore:98} $\sim
L^{-\theta}$.
On the contrary, RSB predicts a nontrivial distribution with a finite weight
in the tail down to $q = 0$, independent of system size.

Earlier
work that studied the nature of the spin-glass state has focused on
the Ising spin
glass,\cite{katzgraber:01,krzakala:00,palassini:00a,marinari:00}
though some work has also been carried out on the gauge
glass model of the vortex glass transition in
superconductors.\cite{katzgraber:01a} Here, we consider a
{\em vector spin-glass}\/ model,
the four-dimensional $XY$ spin glass,
which is known to have a finite
transition temperature\cite{jain:96} $T_c$ with $T_c \simeq 0.95$.
We perform Monte Carlo
simulations for a modest range of sizes down to low temperatures
($T \simeq 0.2 T_c$) using the
parallel tempering Monte Carlo\cite{hukushima:96,marinari:98} technique. 
Our main result is that that $P(0)$ does not appear to
decrease with increasing system size for the range of sizes studied. 

We also look for information on the surface of the large-scale low-energy
excitations by studying the ``link overlap'' defined in
Eq.~(\ref{linkoverlap-xy}) below. The data for this quantity suggests that
the surface may be space filling, i.e., $d_s=d$, as in RSB, though the small
range of sizes precludes us from making a firm statement on this
and a scenario compatible with the DP is also viable in which $d_s < d$.

The layout of the paper is as follows: In Sec.~\ref{model-observables}
we describe the model and the measured observables. We discuss our
equilibration 
tests for the parallel tempering Monte Carlo method for this specific model in
Sec.~\ref{equilibration}. Our results are discussed in Sec.~\ref{results}.
Section \ref{conclusions} summarizes our conclusions
and presents ideas for future work.

\section{Model and Observables}
\label{model-observables}

The $XY$ spin glass consists of two-component spins of unit length 
on a hypercubic lattice in four dimensions with periodic boundary conditions.
The Hamiltonian is 
given by
\begin{equation}
{\mathcal H} = -\sum_{\langle i, j\rangle} J_{ij} {\bf S}_i\cdot {\bf S}_j,
\label{hamiltonian-xy}
\end{equation}
where the sum is over nearest neighbors,
the linear size is $L$, the number of spins is
$N = L^4$, and 
${\bf S}_i \equiv (S_i^x,S_i^y)$ is an $XY$ spin. Since $|{\bf S}_i| = 1$, 
one can parametrize the spins as ${\bf S}_i = [\cos(\phi_i),\sin(\phi_i)]$ 
with $\phi_i \in [0,2\pi]$. The Hamiltonian then transforms to
\begin{equation}
{\mathcal H} = -\sum_{\langle i, j\rangle} J_{ij} \cos(\phi_i - \phi_j).
\label{hamiltonian2-xy}
\end{equation}

The $J_{ij}$ are chosen according 
to a Gaussian distribution with zero mean and standard deviation $J$, i.e.,
\begin{equation}
{\mathcal P}(J_{ij}) = \frac{1}{\sqrt{2\pi} J}
\exp\left[-{ J_{ij}^2 \over 2 J^2 } \right].
\label{disorder-xy}
\end{equation}
Unless otherwise stated we will take $J=1$.

We concentrate on two observables, the spin overlap $q$ and the
link overlap $\ql$.
The (tensor) spin
overlap is defined in terms of the spin configurations of two copies of the system,
denoted by $(1)$ and $(2)$, as follows:
\begin{equation}
q_{\mu \nu} = \frac{1}{N}\sum_{i = 1}^{N} S_{i,\mu}^{(1)} S_{i,\nu}^{(2)}
\qquad, \mu,\nu \in\{x,y\} .
\label{qmunu-xy}
\end{equation}
In analytic work, the spin-glass order parameter is
defined to be the average of
the {\em trace}\/
of $q_{\mu \nu}$. To be precise, for $L \to \infty$,
the order parameter tensor is predicted to
be of the form
\begin{equation}
\pmatrix{
q / 2 & 0 \cr
0   & q / 2 
}.
\label{qtensor}
\end{equation}
However, this implicitly assumes that the symmetry has been
broken by a small field, which 
is inconvenient to implement in numerics, so we
adopt the following equivalent procedure. We apply all possible symmetries
(rotations and reflection)
to one replica and
take the {\em largest}\/
value of the resulting trace. Consider first rotations under
which $q \to q^\prime$ where
\begin{equation}
q^\prime =
\pmatrix{
q^\prime_{xx} & q^\prime_{xy} \cr
q^\prime_{yx} & q^\prime_{yy} } .
\end{equation}
Maximizing 
${\rm Tr}(q^\prime)$ with respect to the
relative rotation angle between the replicas gives
$q_1$, where
\begin{equation}
q_1 
= \sqrt{(q_{xx} + q_{yy})^2 + (q_{yx} - q_{xy})^2}  .
\label{q1-xy}
\end{equation}
The rotation also makes the two off-diagonal pieces equal, i.e., $q^\prime_{xy}
= q^\prime_{yx}$.

We also must consider how the $q_{\mu \nu}$ transform under
reflections of the angles of the spins in one replica, 
$\phi_i \to -\phi_i$.
It is easy to see that under this transformation $q_{1} \rightarrow
q_{2}$ and vice-versa, where
\begin{eqnarray}
q_{2} & = &
\sqrt{ (q^\prime_{xx} - q^\prime_{yy})^2 + (q^\prime_{xy} +
q^\prime_{yx})^2 } \nonumber \\
& = & 
\sqrt{(q_{xx} - q_{yy})^2 + (q_{xy} + q_{yx})^2},
\label{q2-xy}
\end{eqnarray}
where the second line follows after some algebra.
Since the spin-glass order parameter is obtained by maximizing the trace with
respect to all symmetry transformation, it is given by
\begin{equation}
q = {\rm max}\{q_{1},q_{2}\} .
\label{overlap-xy}
\end{equation}
We use the notation $q$, somewhat inconsistently, for the spin-glass order parameter
to conform with notation in other
work.  The spin-glass order parameter function in RSB theory, $P(q)$, is given
by the distribution of $q$ in Eq.~(\ref{overlap-xy}).

We also define the smaller of $q_1$ and $q_2$ by $\qbar$, i.e.,
\begin{equation}
\qbar = {\rm min}\{q_{1},q_{2}\} .
\label{overlap2-xy}
\end{equation}
If the order parameter tensor tends to the form in Eq.~(\ref{qtensor}) for
$L \to \infty$ then $\qbar \to 0$ in this limit. We shall see that our results
support this.

If we are willing to {\em assume}\/ that the form in
Eq.~(\ref{qtensor}) applies in the thermodynamic limit
then we can obtain the-spin glass order parameter
distribution
a little more simply from the quantity
\begin{equation}
Q = \sqrt{q_{xx}^2 + q_{yy}^2 + q_{yx}^2 + q_{xy}^2},
\end{equation}
which is invariant under symmetry transformations. Since
\begin{equation}
2 Q^2 = q^2 + \qbar^2,
\end{equation}
then, if $\qbar \to 0$ for $L \to \infty$,
the distributions of $q$ and $\sqrt{2} Q$ are the
same in this limit.

The link overlap is defined, quite simply, by
\begin{equation}
\ql = \frac{1}{N_b}\sum_{\langle i,j\rangle}
({\bf S}_i^{(1)}\cdot{\bf S}_j^{(1)})
({\bf S}_i^{(2)}\cdot{\bf S}_j^{(2)})\; ,
\label{linkoverlap-xy}
\end{equation}
where $N_b = Nd$ is the number of bonds ($d = 4$ is the space dimension).
Since this is already invariant under global
symmetry operations we do not need to consider the effects of rotations
and reflections as we did for the spin overlap. The link overlap 
can be expressed in terms of spin angles by
\begin{equation}
\ql = \frac{1}{N_b}\sum_{\langle i,j\rangle}
\cos(\phi_i^{(1)} - \phi_j^{(1)})
\cos(\phi_i^{(2)} - \phi_j^{(2)}).
\end{equation}

While a change in $q$ induced by flipping a cluster of spins is
proportional to the {\it volume} of the cluster, $\ql$ changes by an
amount proportional to the {\it surface} of the cluster. The weight in 
$P(q)$ for small $q$ varies as $L^{-\theta^\prime}$, where $\theta^\prime$
was introduced in Sec.~\ref{introduction}. In addition, we expect the
variance of the link overlap to fit to a form ${\rm Var}(\ql) \sim L^{-\mu_l}$
where, as shown in Ref.~\onlinecite{katzgraber:01}, $\mu_l = \theta^\prime
+2(d - d_s)$.

\section{Equilibration}
\label{equilibration}

For the simulations, we use the parallel tempering Monte Carlo
method.\cite{hukushima:96,marinari:98}
In this technique, one simulates identical replicas of the system
at $N_T$ different temperatures, and,
in addition to the usual local moves, one
performs global moves where the temperatures of two replicas (with
adjacent temperatures) are exchanged. This allows us to study
larger systems at lower temperatures than with the conventional Monte Carlo
method.
Since we require two copies at each
temperature to determine the spin and link overlaps, see Eqs.~(\ref{qmunu-xy})
and (\ref{linkoverlap-xy}), we actually simulate $2 N_T$ replicas. 

The lowest temperature has to be far below $T_c \simeq 0.95$ and yet high
enough that a range of sizes can be simulated. We chose the value of 0.2. The highest
temperature has to be such that the system equilibrates very fast, and we
chose $1.498$.
The intermediate temperatures are determined empirically
provided that the acceptance ratios of the
moves interchanging the replicas are larger than about $0.4$ 
and are all roughly equal.

Table~\ref{simparams} lists the parameters of the simulation; $\nsa$ 
(number of samples), $\nsw$ (total number of sweeps performed by each 
set of spins), and $N_T$ (number of temperature values).

\begin{table}
\caption{
\label{simparams}
Parameters of the simulation. $\nsa$ is the
number of samples, i.e.,sets of disorder realizations, $\nsw$ is the total
number
of sweeps simulated for each of the $2 N_T$ replicas for a single sample,
and $N_T$ is the number of temperatures used in the parallel tempering method.
}
\begin{tabular*}{\columnwidth}{@{\extracolsep{\fill}} c r r l }
\hline                                                                                    
\hline
$L$  &  $\nsa$  & $\nsw$ & $N_T$  \\ 
\hline
3 & $1 \times 10^4$ & $3.0 \times 10^4$ &   39 \\
4 & $2 \times 10^3$ & $4.0 \times 10^5$ &   39 \\
5 & $1 \times 10^3$ & $2.0 \times 10^6$ &   39 \\
\hline
\hline
\end{tabular*}
\end{table}

\begin{figure}
\centerline{\epsfxsize=\columnwidth \epsfbox{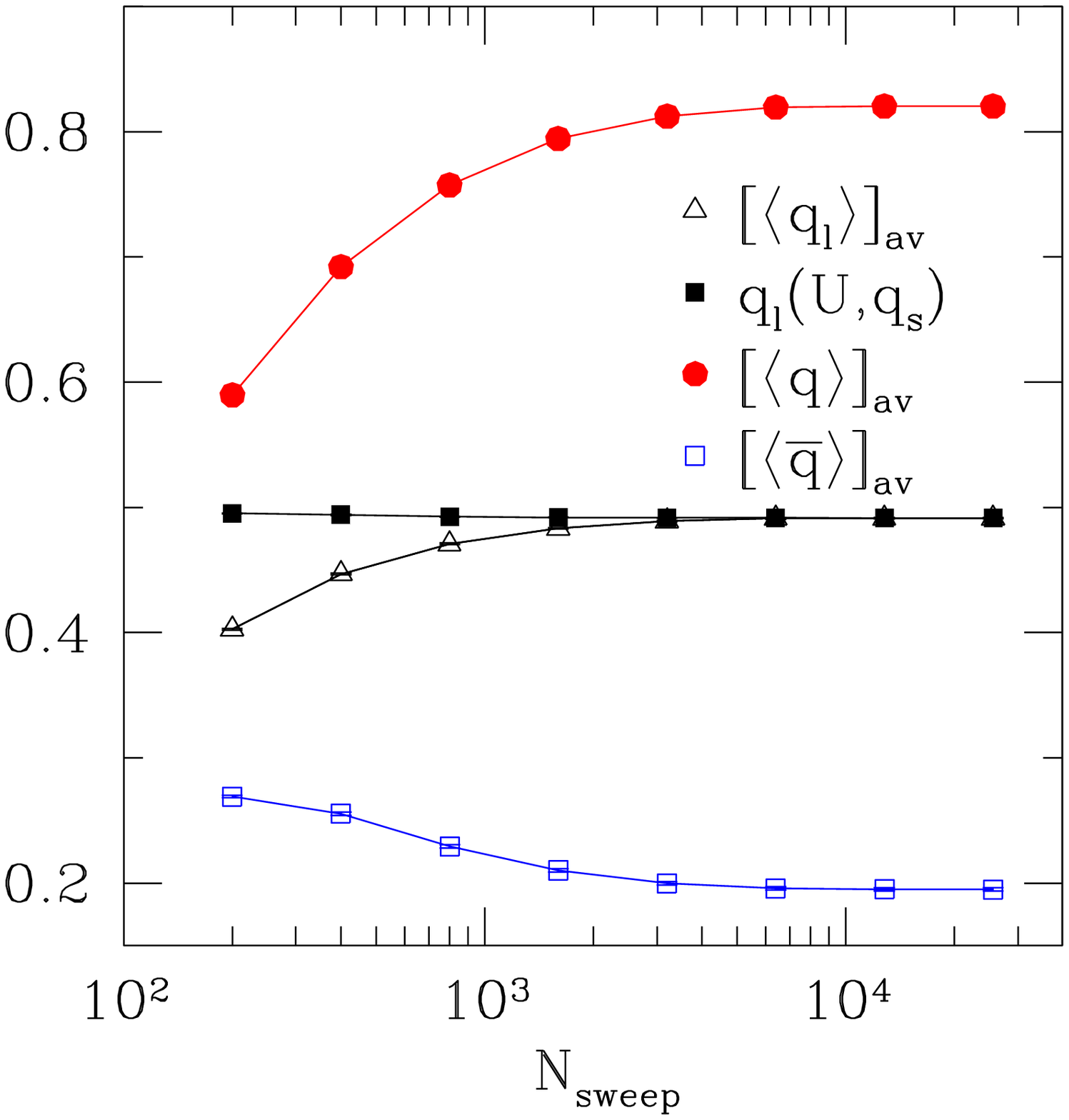}}
\caption{
A plot of
$[\langle \ql\rangle ]_{\rm av}$ (the link overlap), $\ql(U,q_s)$ 
defined to be the RHS of Eq.~(\ref{equilrel}),
$[\langle q\rangle ]_{\rm av}$
the spin overlap, and
$[\langle \qbar\rangle ]_{\rm av}$ defined in Eq.~(\ref{overlap2-xy}),
as a function of Monte Carlo sweeps $\nsw$ for each 
replica, averaged over the last half of the sweeps. For 
equilibration, $[\langle \ql\rangle ]_{\rm av}$ and $\ql(U,q_s)$
should agree. The two sets of data approach each other
from opposite directions and,
once converged, do
not seem to change at longer times, indicating that the system
is equilibrated. The
data for $[\langle q\rangle ]_{\rm av}$ and $[\langle \qbar\rangle
]_{\rm av}$ show that they too have equilibrated in roughly the same
equilibration time. While not shown here, data for higher moments of the
different observables have the same equilibration time as the link overlap
$[\langle \ql\rangle ]_{\rm av}$.  (Data for $L = 3$, $T = 0.2$, and 3230
samples).
}
\label{equil}
\end{figure}

It is important to ensure that the system is equilibrated.
However,
the equilibration test proposed by Bhatt and Young\cite{bhatt:85} does not 
work with parallel tempering Monte Carlo because the temperature of each
replica
does not stay constant throughout the simulation. Here we use the method 
introduced by Katzgraber {\em et al}.\cite{katzgraber:01} 
for short-range spin glasses with a Gaussian distribution of exchange
interactions
that relates the average energy to the link overlap. By performing an
integration 
by parts with respect to $J_{ij}$ of the average energy
$U \equiv [\langle {\cal H} \rangle ]_{\rm av} \ (\le 0)$, we obtain
\begin{equation}
 [\langle \ql \rangle ]_{\rm av} = q_s - {2 \over z} { T|U| \over J^2},
\label{equilrel}
\end{equation}
where $z$ is the number of nearest neighbors, 
$\langle \cdots \rangle$ denotes a thermal average, and $[ \cdots ]_{\rm av}$
denotes an average over the disorder.
The quantity $q_s$ is given by
\begin{equation}
q_s = \frac{1}{N_b}\sum_{\langle i,j\rangle}
[ \langle ({\bf S}_i\cdot{\bf S}_j) ^2 \rangle ]_{\rm av},
\label{qs}
\end{equation}
where
the sum is over pairs of
neighboring spins. The simulation is started with randomly chosen spins so
that
all replicas are uncorrelated. This will have the effect that both sides of
Eq.~(\ref{equilrel}) are approached from opposite directions. Once they
agree,
the system is in equilibrium as can be seen in Fig.~\ref{equil} for
$T = 0.2$ (to be compared with $T_c \approx 0.95$),\cite{jain:96}
the lowest temperature simulated, and for $L = 3$. We show data for the
smallest size since it allows us to generate more samples for longer
equilibration times to better illustrate the method. For larger system sizes
we stop the simulation, once the data for $ [\langle \ql \rangle ]_{\rm av}$
and the right-hand side (RHS) of Eq.~(\ref{equilrel}) agree.

Because the $XY$ spin glass
has a vector order parameter symmetry, we discretize the 
angles of the spins to $N_{\phi} = 512$ to speed up the simulation.
This number is large enough to avoid any crossover effects to other models as 
discussed by Cieplak {\em et al}.\cite{cieplak:92}.
To ensure a reasonable acceptance ratio for single-spin
Monte Carlo moves, we choose the proposed new angle for a spin
within an acceptance window about the current angle, where
the size of the window is proportional to the temperature $T$. By tuning a 
numerical prefactor, we ensure the acceptance ratios 
for these local moves are not smaller than 0.4 for each system size at the
lowest temperature simulated.

\section{Results}
\label{results}

\begin{figure}
\centerline{\epsfxsize=\columnwidth \epsfbox{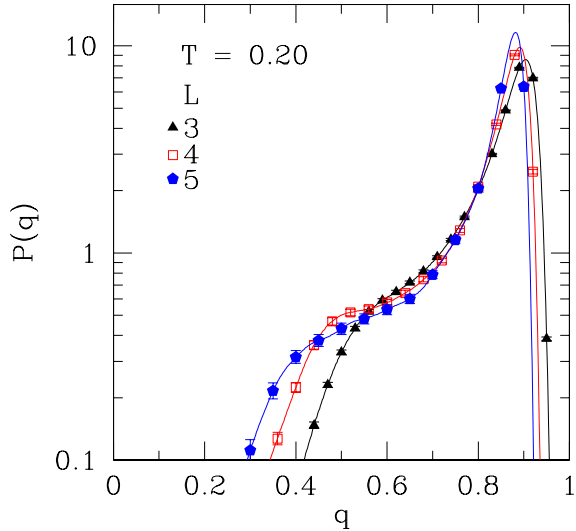}}
\caption{
Data for the spin overlap distribution $P(q)$ at temperature $T = 0.20$ 
for different system sizes. Note the logarithmic vertical scale. The lines go
through all the data points but, for clarity, only some of the data points are
shown.
}
\label{pqd-0.20-xy}
\end{figure}
\begin{figure}
\centerline{\epsfxsize=\columnwidth \epsfbox{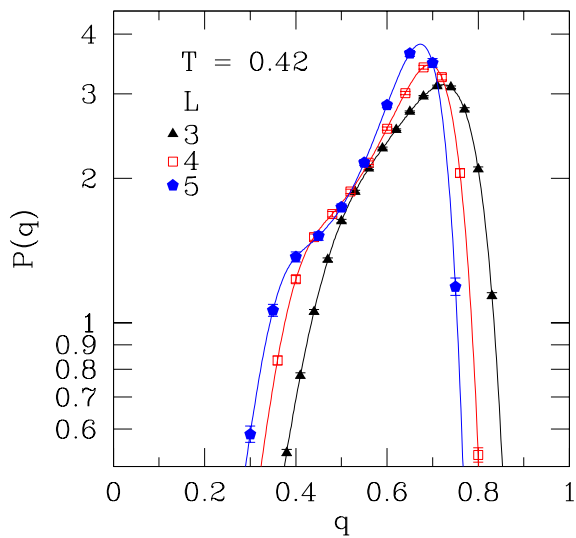}}
\caption{
Same as Fig.~\ref{pqd-0.20-xy} but at temperature $T = 0.42$.
}
\label{pqd-0.42-xy}
\end{figure}

Figures~\ref{pqd-0.20-xy} and \ref{pqd-0.42-xy} show data for $P(q)$ for 
$T = 0.20$ and $0.42$, respectively. In both cases we see a peak at large $q$
and a tail for smaller $q$ that does not extend to $q=0$. However, it is not
surprising that there is a ``hole'' at small $q$ since $q$ is defined to be
the maximum of $q_1$ and $q_2$. If $\qbar \equiv \min\{q_1, q_2\}$ tends to
zero at large $L$, which is expected as discussed above, then, in RSB theory,
the tail would
extend to smaller values of $q$ for larger $L$ while maintaining the same
height. Looking at
Figs.~\ref{pqd-0.20-xy} and \ref{pqd-0.42-xy}, this seems to be the case, at
least for the range of sizes that we have been able to study.

\begin{figure}
\centerline{\epsfxsize=\columnwidth \epsfbox{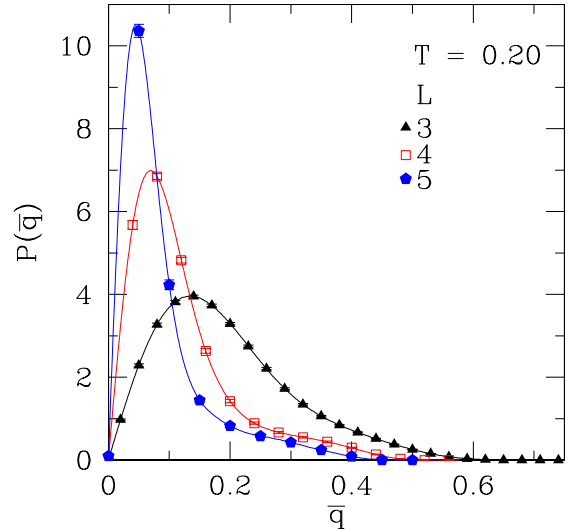}}
\caption[Distribution of $\qbar$ at $T = 0.20$ in the 4D $XY$ spin-glass]
{Data for the overlap distribution $P(\qbar)$ at temperature $T =
0.20$
for different system sizes. The weight in the distribution tends towards
$\qbar = 0$ for increasing $L$.
}
\label{pqo-0.20-xy}
\end{figure}

\begin{figure}
\centerline{\epsfxsize=\columnwidth \epsfbox{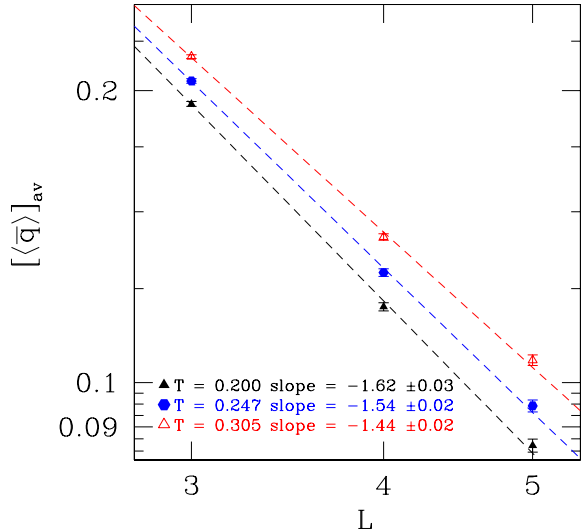}}
\caption[
Mean of $\qbar$ as a function of size in the 4D $XY$
spin-glass]
{
Log-log plot of $[\langle \qbar \rangle ]_{\rm av}$ as a function of system size 
$L$ at several temperatures.
}
\label{varqo-xy}
\end{figure}

In Fig.~\ref{pqo-0.20-xy}
we show data for 
$P(\qbar)$ at $T = 0.20$.
As expected, the 
distributions seem to collapse to zero for 
increasing system size. 
Figure~\ref{varqo-xy} shows the variation of the mean of $\qbar$ with $L$ on a
log-log plot. The data have been fitted to straight lines with slopes shown.
The quality of the fits\cite{press:95}
is only moderate; $Q = 0.06, 0.09$, and $0.04$
for $T=0.200, 0.247$, and $0.305$, respectively. Given the rather small range
of sizes, and hence the likelihood of systematic corrections to scaling,
we feel that the data are consistent with $ [\langle \qbar \rangle ]_{\rm av}
\to 0$ for $L \to \infty$. Since $\qbar \ge 0$, if $ [\langle \qbar \rangle
]_{\rm av} = 0$ then the whole distribution collapses to $\qbar = 0$.

\begin{figure}
\centerline{\epsfxsize=\columnwidth \epsfbox{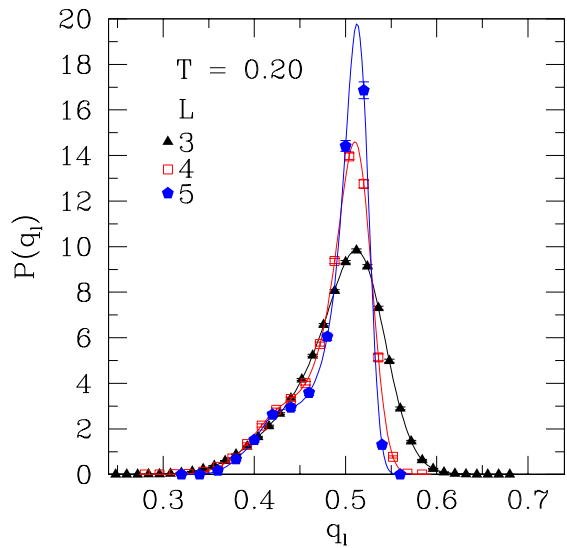}}
\caption[Distribution of the link overlap in the 4D $XY$ spin-glass at $T =
0.20$]
{
The distribution of the link overlap at $T = 0.20$ for different
sizes.
}
\label{pqb-0.20-xy}
\end{figure}

\begin{figure}
\centerline{\epsfxsize=\columnwidth \epsfbox{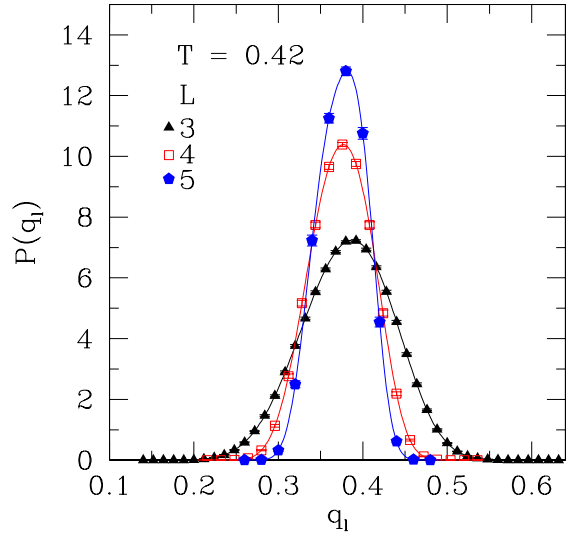}}
\caption[Distribution of the link overlap in the 4D $XY$ spin-glass at $T =
0.46$]
{
Same as Fig.~\ref{pqb-0.20-xy} but at temperature $T = 0.42$.
}
\label{pqb-0.42-xy}
\end{figure}

\begin{figure}
\centerline{\epsfxsize=\columnwidth \epsfbox{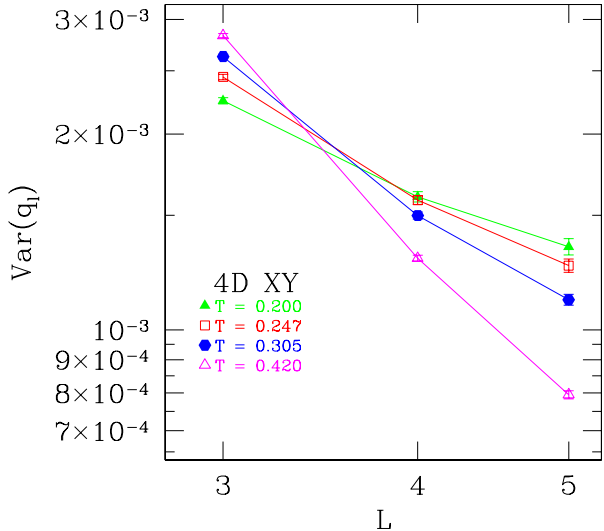}}
\caption{
Log-log plot of the variance of $\ql$ as a function of system size 
$L$ at several temperatures.
}
\label{varqb-xy}
\end{figure}

Lastly we present in Figs.~\ref{pqb-0.20-xy} and \ref{pqb-0.42-xy}
results for the distribution of the link overlap 
$P(\ql)$. There is a pronounced peak at large $\ql$ values as well 
as the hint of a shoulder for smaller values in the $T=0.20$ data.
The width of
the distribution decreases with increasing system size. This is demonstrated
in Fig.~\ref{varqb-xy}, which shows the variance of $\ql$
against system size $L$ for several low temperatures.

There is some curvature in the data for $\varql$, so first we attempt a
three-parameter fit of the form
\begin{equation}
\varql = a + bL^{-c}\; ,
\label{3params}
\end{equation}
finding small but finite values for $a$, see Table~\ref{table1.xy}.
As we have the same number of data points as variables, 
we cannot assign fitting probabilities to
the fits. 
We also attempt a power-law fit of the form 
\begin{equation}
\varql = dL^{-\mu_l} ,
\label{2params}
\end{equation}
see Table~\ref{table2.xy}. However, the quality of the
fits is poor as shown by the fitting
probabilities\cite{press:95} $Q$.
The effective
exponent $\ml$ is found to vary with temperature. 
Extrapolating to $T = 0$, 
we obtain $\ml \equiv \theta^\prime
+ 2(d - d_s) =  0.294 \pm 0.073$.
If we assume that $\theta^\prime \approx 0$, this gives
$d - d_s = 0.147 \pm 0.036$.

\begin{table}
\caption{
\label{table1.xy}
Fits for ${\rm Var}(\ql)$.
Fit parameters for the fit in Eq.~(\ref{3params})
for different temperatures.
We cannot quote fitting
probabilities since we have the same number of data points as variables.
}
\begin{tabular*}{\columnwidth}{@{\extracolsep{\fill}} c r r l }
\hline   
\hline
$T$  &  $a$ & $b$  & $c$ \\
\hline
0.200 & $0.00100$ & 0.0205 & $2.55$ \\
0.247 & $0.00087$ & 0.0328 & $2.76$ \\
0.305 & $0.00073$ & 0.0611 & $3.16$ \\
0.420 & $0.00036$ & 0.1044 & $3.40$ \\
\hline
\hline
\end{tabular*}
\end{table}

\begin{table}
\caption{
\label{table2.xy}
Fit parameters for the fit in Eq.~(\ref{2params})
for different temperatures.
Note that the fit probabilities $Q$ are small.
}
\begin{tabular*}{\columnwidth}{@{\extracolsep{\fill}} c r r l }
\hline   
\hline
$T$  &  $d$ & $\mu_l$ & $Q$ \\
\hline
0.200 & $-4.92 \pm 0.06$ & $1.07\pm 0.05$ & $5.0 \times 10^{-2}$ \\
0.247 & $-4.50 \pm 0.05$ & $1.38\pm 0.04$ & $3.6 \times 10^{-3}$ \\
0.305 & $-3.99 \pm 0.04$ & $1.77\pm 0.03$ & $2.9 \times 10^{-6}$ \\
0.420 & $-3.06 \pm 0.04$ & $2.56\pm 0.03$ & $6.0 \times 10^{-8}$ \\
\hline
\hline
\end{tabular*}
\end{table}

\section{Conclusions}
\label{conclusions}

To conclude, we have studied the low-temperature properties of the
four-dimensional $XY$ spin glass at
low temperatures. 
Our main result is that
the order parameter distribution $P(q)$ has, in addition to a
peak, a tail that seems to extend for smaller values of $q$, and
whose height seems to persist, as the system size increases, see
Figs.~\ref{pqd-0.20-xy} and \ref{pqd-0.42-xy}. This
interpretation of the data is 
compatible with the RSB picture or the KMPY scenario. However, the range of
lattice sizes is very small, so it is not clear if this interpretation would
persist to large sizes. Unfortunately, it is currently not feasible to study
much larger sizes in equilibrium, because relaxation times are too long.
Nonetheless, we feel that results on rather small equilibrated samples are of
interest in their own right
for the following reason: In any experiment, a sample is not fully
equilibrated at low temperatures, but is rather only equilibrated up to some
finite length scale, which only increases slowly with increasing measurement
time. Thus a complete understanding will require a {\em nonequilibrium
theory}\/, but a component of this is likely to be a theory of equilibrium on
{\em finite}\/ scales where local equilibrium has been achieved.

We have also studied the link overlap $\ql$. The variance of $\ql$ decreases
with increasing $L$ but we are unable to ascertain whether it tends to zero in
the thermodynamic limit, and hence we are unable to determine whether or not
the surface is space filling.  
 
In future work, it would be useful to look more carefully at the nature of the
large-scale low-energy excitations to see whether they correspond to gradual
orientations in the spin directions or whether vortices play a role.

\begin{acknowledgments}
This work was supported by the National Science Foundation under grant 
No.~DMR 0086287.
We are grateful to D.~A.~Huse for correspondence about the possible role of
vortices.
The numerical calculations were made possible by use of the UCSC 
Physics graduate computing cluster funded by the Department of
Education Graduate Assistance in the Areas of National Need program.
We would also like to thank the University of New Mexico for access to their
Albuquerque High Performance Computing Center. This work utilized the
UNM-Truchas Linux Clusters.
\end{acknowledgments}

\end{document}